\documentclass[aps,reprint,superscriptaddress,showpacs]{revtex4-1}
\usepackage{microtype, textcomp}
\usepackage[abbreviations = false, detect-all, list-units = single]{siunitx}
\usepackage[inline]{enumitem}
\usepackage{graphicx}

\usepackage{amsmath, amssymb}
\usepackage[english]{babel}

\usepackage{xspace}

\newcommand{\one}{\num{e14}\xspace}
\newcommand{\onepointosixteen}{\num{0.5e14}\xspace}
\newcommand{\onepointone}{\num{e13}\xspace}

\begin{document}

\title{Diffusion of light in semitransparent media}

\author{Lorenzo Pattelli}
\email[]{pattelli@lens.unifi.it}
\affiliation{European Laboratory for Non-linear Spectroscopy (LENS), Universit\`{a} di Firenze, 50019 Sesto Fiorentino (FI), Italy}

\author{Giacomo Mazzamuto}
\affiliation{European Laboratory for Non-linear Spectroscopy (LENS), Universit\`{a} di Firenze, 50019 Sesto Fiorentino (FI), Italy}
\affiliation{CNR-INO, Istituto Nazionale di Ottica, Via Carrara 1, 50019 Sesto Fiorentino (FI), Italy}

\author{Diederik S. Wiersma}
\affiliation{European Laboratory for Non-linear Spectroscopy (LENS), Universit\`{a} di Firenze, 50019 Sesto Fiorentino (FI), Italy}
\affiliation{Department of Physics, Universit\`{a} di Firenze, 50019 Sesto Fiorentino (FI), Italy}

\author{Costanza Toninelli}
\affiliation{European Laboratory for Non-linear Spectroscopy (LENS), Universit\`{a} di Firenze, 50019 Sesto Fiorentino (FI), Italy}
\affiliation{CNR-INO, Istituto Nazionale di Ottica, Via Carrara 1, 50019 Sesto Fiorentino (FI), Italy}
\affiliation{QSTAR, Largo Enrico Fermi 2, 50125 Firenze, Italy}

\date{\today}

\begin{abstract}
Light diffusion is usually associated with thick, opaque media. Indeed, multiple scattering is necessary for the onset of the diffusive regime and such condition is generally not met in almost transparent media. Nonetheless, at long enough times, transport in an infinite thin slab is still determined by a multiple scattering process whose complete characterization is lacking. In this paper we show that, after a short transient, the mean square width of the transmitted intensity still exhibits a simple linear increase with time as predicted by diffusion theory, even at optical thickness as low as one. Interestingly, such linear growth is predicted not to depend neither on the slab thickness nor on its refractive index contrast, yet the accuracy of this simple approximation in the ballistic-to-diffusive regime hasn't been investigated so far.
By means of Monte Carlo simulations, we find clear evidence that boundary conditions play an active role in redefining the very asymptotic value of the diffusion coefficient by directly modifying the statistical distributions underlying light transport in such media. In this respect, we demonstrate the need to distinguish between a set of \emph{intrinsic} and \emph{effective} transport parameters, whose relation and interplay with boundary conditions remains to be fully understood.
\end{abstract}


\maketitle

\section{\label{sec:intro}Introduction}
The modeling of light transport in disordered media represents an intensive research field both for its fundamental interest and for its applications. In this ubiquitous class of media, because of the stochastic nature of multiple scattering and absorption, light transport can effectively be described as a random walk of energy packets \cite{chandrasekhar1950radiative, ishimaru1978wave}. 
Whenever an unbounded or thick geometry can be assumed, the diffusion approximation (DA) provides an incredibly robust and useful simplification of such problem since it casts a complete set of analytic predictions on all transport observables, also in the slab geometry \cite{contini1997photon}. Their validity relies on the assumption of almost isotropic radiance, which is underlying in the diffusion approximation but progressively breaks down if the characteristic scattering length becomes comparable to absorption or the thickness of the slab.
In the latter case a significant fraction of light is usually able to go through the sample without undergoing any scattering event and the slab becomes semitransparent.
This configuration is particularly relevant in many applications, ranging from coatings to photovoltaic and biomedical, and typically requires a refined analysis to overcome the limitations of the diffusive approximation.
In this case Monte Carlo (MC) simulations are typically considered as the gold standard method to solve the radiative transfer problem, in that they allow a direct implementation of the random walk model with accurate boundary and geometric conditions \cite{wang1995mcml}.
With the continuous growth of computing power and parallelization capabilities, Monte Carlo simulations are gaining momentum as a convenient solver both for the forward and inverse light transport problem. Indeed, by resorting to MC simulations, novel insights are continuously unveiled on the peculiar physics that governs light transport in complex media such as heterogeneous or anisotropic materials \cite{svensson2013holey, alerstam2014anisotropic}.
Conversely, we show that our comprehension of way simpler systems such a homogeneous, isotropic single slab is still incomplete.

To this purpose we investigate by means of Monte Carlo simulations what is usually referred to as the diffusive-to-ballistic transition in the thin slab geometry, with a particular focus on \emph{transverse} transport. Indeed the thin slab geometry is typically associated with a ballistic transport regime and consequently to the inapplicability of diffusion theory. Instead, the former does not necessarily imply the latter, which becomes clear when addressing in-plane transport where a deeply multiple scattering regime can indeed occur \cite{leonetti2011measurement}. In an infinitely extended slab, transverse propagation is unbounded and will eventually become diffusive at sufficiently long times. Nonetheless, even if asymptotic, we will show that this diffusive regime is not the one predicted by diffusion theory based on the intrinsic transport properties of the investigated medium. In this respect our findings notably show that, for example, even a minute tuning in the refractive index contrast allows to sensibly modify the long time behavior of semitransparent slabs, which might therefore be still functionalized in terms of scattering properties in spite of their low optical thickness.

The paper is organized as follows: in Section \ref{sec:msw} it is shown that a typical transverse diffusion signature such as the linear growth of the mean square width (MSW) is still clearly visible after a short transient even at an optical thickness as low as 1. In Section \ref{sec:deviations} we find and characterize deviations of both the expected mean square width slope and transmittance decay lifetime from their expected values according to the diffusive approximation, discussing also experimental implications. In Section \ref{sec:discussion} we describe the origin of the observed deviations in terms of a statistical analysis of simulated transport, unveiling the peculiar features of a diffusion process which remarkably differs from the one expected by standard diffusion theory.

\section{\label{sec:msw}Transverse transport in a slab geometry}
\begin{figure*}
\centering
\includegraphics{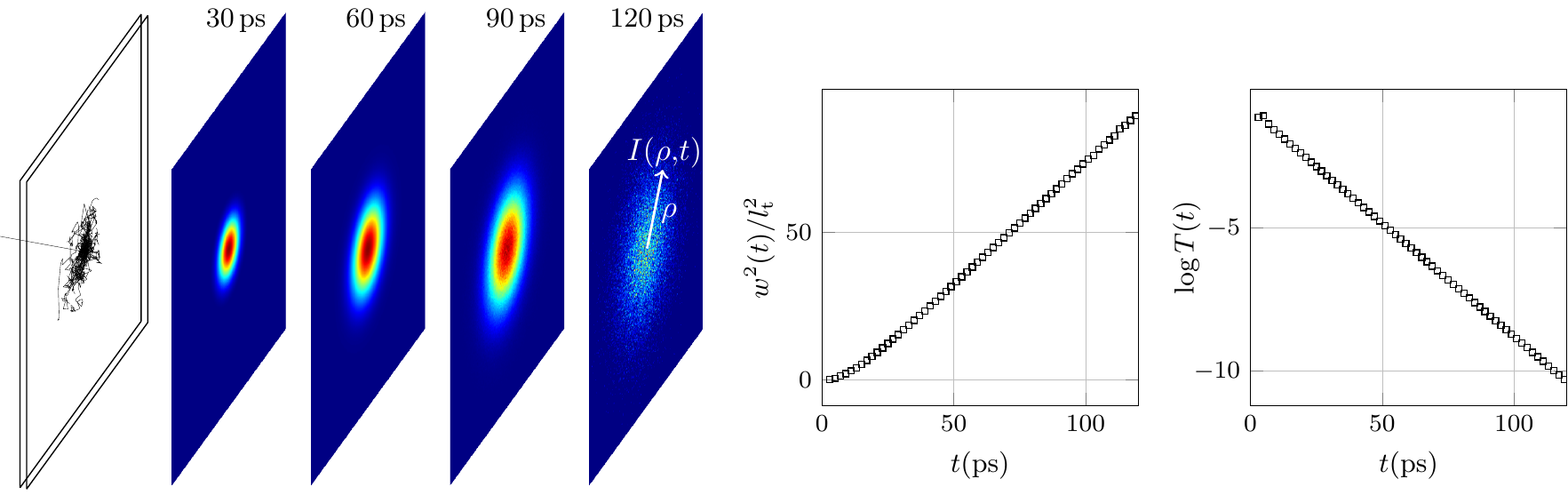}
\caption{Sketch of the investigated configuration. An infinite slab is illuminated by a pencil beam $\delta (t)$ pulse. Transmitted light is collected at different times and positions. A few (longer) trajectories and normalized transmitted intensities are presented for illustrative purposes in the case of an optically thin, index matched slab with thickness and scattering mean free path equal to \SI{1}{\milli\meter} (i.e.\ $\text{optical thickness} = \num{1}$), shown to scale. The scattering anisotropy $g$ is set to \num{0}. An approximately Gaussian profile is transmitted at each time slice, whose mean square width and integrated intensity respectively grows linearly and falls exponentially.}
\label{fig:cartoon}
\end{figure*}
The slab geometry (see Fig.\ \ref{fig:cartoon}) is a very simple yet relevant model in a wide range of applications. As such, transport properties in this configuration have been extensively addressed in the literature and tested against diffusion theory, but mainly focusing on axial rather than lateral migration \cite{yoo1990does, yoo1990time, lemieux1998diffusing, zhang1999wave, zhang2002wave, elaloufi2004diffusive, svensson2013exploiting}. This is surprising since the theory fully describes transverse transport as well, providing the simplest relations linking experimental observables with the transport properties of the sample. According to the standard diffusion approximation, the instantaneous spatial profile emerging from a slab characterized by a diffusion coefficient $D$ is always expected to be a Gaussian distribution $I(\rho, t) \propto \exp (-\rho^2 / w^2 (t))$ with a mean square width growing as $w^2 (t) = 4Dt$. Such width represents the variance, which is more generally defined for an arbitrary distribution $I(\rho, t)$ as
\begin{equation}
w^2(t) = \frac{\int \rho^2 I(\rho, t) \, \mathrm{d}\rho}{\int I(\rho, t) \, \mathrm{d}\rho}.
\label{eq:MSW}
\end{equation}
The mean square width value is most remarkably independent of absorption (which cancels out exactly at any time \cite{cherroret2010transverse}) and, according to the simple diffusive model, independent of both the slab thickness and its refractive index contrast with the surrounding environment. This stands in great contrast with any other expression available for the slab geometry in the diffusive approximation, where the correct assessment of absorption, thickness and extrapolated boundary conditions play a critical role in correctly retrieving the optical properties of the sample \cite{zhu1991internal, bouchard2010reference}, thus representing a largely appealing quantity to be experimentally investigated \cite{pattelli2015direct}. Notably, even at an optical thickness as low as \num{1}, a perfectly linear mean square width growth is permanently recovered soon after a short superlinear transient, as Figure \ref{fig:cartoon} illustrates for an index matched sample with unitary refractive index. As shown in the following, a qualitatively similar behavior is actually common to a broad range of optical parameters, which further makes the mean square width a valuable experimental observable also at low optical thicknesses.

\section{\label{sec:deviations}Mean square width expansion}
\begin{figure}
\centering
\includegraphics{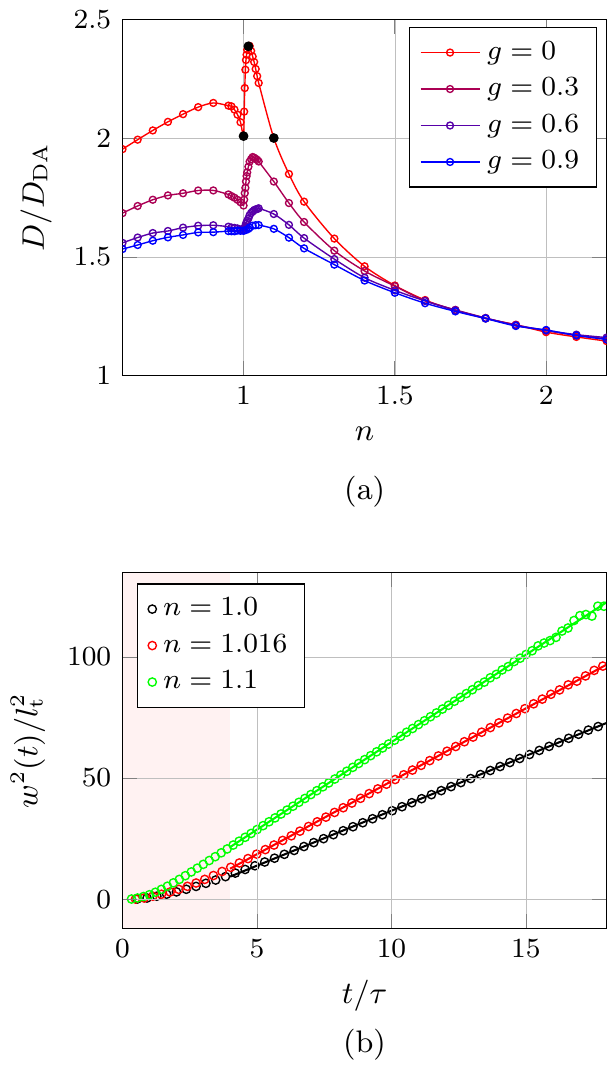}
\caption{(a) Cross-cuts along different values of $g$ of the simulated hyper-surface dataset \cite{mazzamuto2015breakdown}, taken at $l_\text{t} = L_0 = \SI{1}{\milli\meter}$ ($\text{OT} = 1$). For each point, the MSW slope has been evaluated from a simulation of \num{e10} energy packets. The introduction of a small amount of boundary reflection appears to enhance in-plane diffusion remarkably, which also affects $g > 0$ and $\text{OT} > 1$ samples \cite{mazzamuto2015breakdown}. A subset of parameters (black filled dots) where the effect is more dramatic is further investigated in the following, based on a much larger statistics. (b) MSW growth for the three different configurations highlighted in the upper panel. Linear fits are performed excluding $t<4\tau$.}
\label{fig:OT1_plane}
\end{figure}

The validity range of the simple $w^2 (t) = 4Dt$ prediction has never been tested, to our knowledge. A comprehensive Monte Carlo investigation, whose complete set of results are reported in a related work \cite{mazzamuto2015breakdown}, revealed the presence of non-trivial discrepancies which challenge the established framework linking radiative transfer to diffusion theory. For a single slab geometry with fixed thickness $L_0=\SI{1000}{\micro\metre}$ we extracted the transmittance mean square width growth rate from a large set of different simulations in the ($n$, $g$, $\text{OT}^{-1}$) parameter space, where $n=n_\text{in}/n_\text{out}$ is the refractive index contrast, $g$ is the scattering anisotropy factor and $\text{OT}^{-1} = l_\text{t}/L_0$ --- a dimensionless measure for the transport mean free path $l_\text{t}$ --- is the inverse of the optical thickness. For each simulation we estimate $D$ by performing a linear fit on the mean square width $w^2(t>4\tau)$ as a function of time, where $\tau$ is the simulated transmittance decay lifetime and $4\tau$ is selected as a lower limit so to always exclude early-time energy packets transmitted before the onset of the diffusive regime (see Figures \ref{fig:OT1_plane}a and b). The obtained values for $D$ are compared to the ones expected from the diffusion approximation assuming $D_\text{DA} = v l_\text{t} / 3$. Figure \ref{fig:OT1_plane}a shows a subset of simulated mean square width slopes obtained from samples with $l_\text{t} = L$.

As can be seen, high deviations are obviously to be found at such a low optical thickness, especially in the proximity of $n=1$, which is unexpected since this condition is sometimes quoted as a safer configuration for the diffusion approximation \cite{svensson2013exploiting}. It is also worth noting that the mean square width slope of simulated data is always, over the whole set of parameters \cite{mazzamuto2015breakdown}, equal to or greater than the value expected by standard diffusion with the ratio approaching 1 for increasing optical thicknesses, thus recovering the diffusive prediction as expected.
This general trend can be intuitively explained considering the $d$-dimensional modeling of diffusion as a random walk process, which, given any arbitrary step length distribution $P(l)$ with finite moments $\langle l \rangle$ and $\langle l^2 \rangle$, predicts a mean square $d$-dimensional displacement growing as $2dDt$ with
\begin{equation}
D = \frac{1}{2d} v \frac{\langle l^2 \rangle}{\langle l \rangle} = \frac{1}{d} \, v \, l_\text{t}
\label{eq:D(d)}
\end{equation}
where the last equality holds for an exponential step length distribution with average step length $l_\text{t}$ \cite{svensson2013holey}. Indeed, as the optical thickness of the simulated slab decreases, transport occurs in an increasingly planar geometry. Hence, as suggested by eq.~\eqref{eq:D(d)}, the effective diffusion coefficient $D$ as inferred from the mean square width slope might be up to $3/2$ times higher than its bulk nominal value. The perceived environment dimensionality is also clearly affected by the refractive index contrast. Near $n=1$, where deviations are more relevant, any late-time surviving energy packet will have performed an almost planar trajectory, more similar to a purely 2-dimensional walk. On the contrary, with increased reflections at boundaries, trajectories are allowed to fold back into the sample thus perceiving a more 3-dimensional environment.

A closer look on the data reveals an unexpected feature occurring in the close proximity of $n=1$. A set of $g$-cross-cuts in the plane $l_\text{t}/L_0 = 1$ (Figure \ref{fig:OT1_plane}a) reveals a sharp modulation of the mean square width slope across unitary index contrast. In particular, contrarily to what one would expect from the above discussion, the mean square width expansion exhibits a local \emph{minimum} at $n=1$ rather than a maximum. In its close proximity, diffusion is asymmetrically enhanced reaching an absolute maximum around $n=1.016$ for $g=0$.
These features arise from a subtle interplay between the optical properties of the sample, as we will discuss in the following section.

\begin{figure}
\centering
\includegraphics{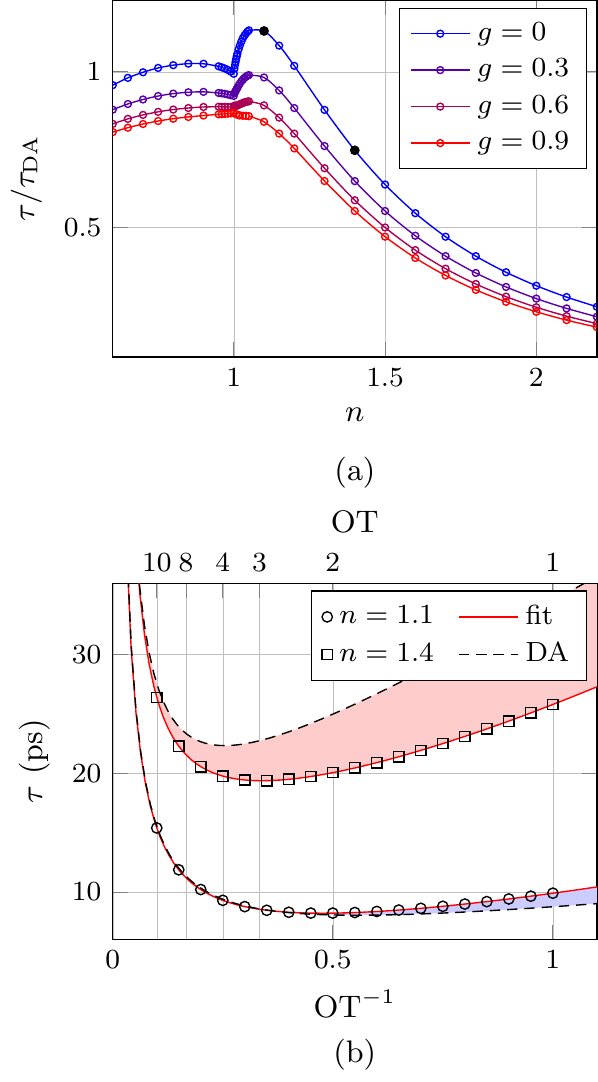}
\caption{
\looseness=-1
(a) Detailed view of the $n$ dependence of the lifetime deviation at $\text{OT} = 1$. Especially for lower $g$ factors, the ratio can clearly go above 1 for specific contrasts. The dependence of $\tau$ on OT for values highlighted as black filled circles is shown in panel (b), along with the DA prediction and a higher order fit.}
\label{fig:OT1_lifetimes}
\end{figure}

Before any further discussion on the mean square width expansion, it is anyway worth noting that a qualitatively similar behavior is mirrored in the lifetime relative deviations from the diffusive prediction, and as such this effect is not strictly limited to transverse transport. This is of particular interest since lifetime measurements have long been experimentally accessible and exploited to estimate the diffusion coefficient via the relation (in case of a non-absorbing medium)
\begin{equation}
\tau_\text{DA} = \frac{L_\text{eff}^2}{\pi^2 D_\text{DA}},
\label{eq:tauDA}
\end{equation}
where $L_\text{eff} = L + 2z_\text{e}$ is the effective thickness of the medium and $z_\text{e}$ represents the extrapolated boundary conditions for a given refractive index contrast.
The similarity in the dependence on $n$ is particularly evident when plotting the same cross-cuts as for the mean square width slope relative deviations (Figure \ref{fig:OT1_lifetimes}a). Two main features are worth commenting. Firstly, it is evident that the lifetime ratio can deviate more significantly from the diffusion approximation than the mean square width slope. Substantial deviations from the lifetime prediction are indeed known to persist also at higher optical thicknesses, especially for $n>1$ despite the correct extrapolated boundary conditions being applied \cite{elaloufi2002time, zhang2002wave}. As shown by our complete set of simulations \cite{mazzamuto2015breakdown}, the mean square width prediction error at $\text{OT} = \num{10}$ does not exceed $\approx \SI{2}{\percent}$ for each refractive index contrast and scattering anisotropy probed, while the same deviations for the lifetime can exceed \SI{20}{\percent} for $n>1$.
The second relevant difference regards the fact that the $\tau/\tau_\text{DA}$ ratio can evidently take values both above and below \num{1}, depending subtly on the scattering anisotropy and the refractive index contrast of the sample. This observation might provide a clearer picture on why a diffusion-based retrieval of the diffusion coefficient from a lifetime measurement is sometimes regarded as a poor estimation, since it can lead both to over- or underestimated values (Figure \ref{fig:OT1_lifetimes}a). This is further illustrated in Figure \ref{fig:OT1_lifetimes}b for a couple of representative cases exhibiting opposite deviations that can persist even at higher optical thicknesses.
To date indeed, experimental data and theoretical predictions are inconsistent. While the former bring generally evidences suggesting a decreasing diffusion coefficient with decreasing thickness \cite{kop1997observation, rivas2001static}, the latter have so far mainly provided arguments in favor of the opposite behavior \cite{ramakrishna1999diffusion, gopal2001photon, zhang2002wave, elaloufi2002time}.
In this respect it is worth noting that our simulations suggest that there is a region in the parameter space where the $\tau/\tau_\text{DA}$ ratio could exceed 1, which can lead to the observed decreasing diffusion coefficient with decreasing thickness. The analysis on the decay lifetimes confirms the importance of an accurate and precise modeling of the index contrast, which we think has been often overlooked, for example by considering a symmetric averaged contrast to model asymmetric experimental configurations \cite{ramakrishna1999diffusion, zhang2002wave, elaloufi2002time}.

It is worth noting that a better modeling of the $\tau(l_\text{t})$ dependence can be envisioned, given its inappropriateness at low optical thicknesses. In the standard diffusion approximation, the lifetime equation can be written as
\begin{equation}
\tau = \frac{(L+2z_\text{e})^2}{\pi^2 D} = p_0 + p_{+1} l_\text{t} + p_{-1} l_\text{t}^{-1}
\label{eq:hiord_tau1}
\end{equation}
where $p_0 = 8AL/\pi^2 v$, $p_{+1} = 16A^2/3\pi^2 v$, $p_{-1}= 3L^2/\pi^2 v$ and $A = A(n)$ is the correction factor accounting for internal reflections. By including also higher order terms as $p_{+2} l_\text{t}^{2}$ and $p_{-2} l_\text{t}^{-2}$ it is indeed possible to perfectly reproduce each simulated value (Fig.\ \ref{fig:OT1_lifetimes}b, red curves) on a broad parameter space, while preserving important physical properties of the lifetime dependence such as its divergence with $\text{OT} \rightarrow \infty$, which would not be guaranteed by a generic polynomial fit.

\section{\label{sec:discussion}Discussion}
In order to explain the origin of the features shown above for the mean square width dependence on $n$, we now focus on three significant sets of parameters (highlighted as black filled circles in Figure \ref{fig:OT1_plane}a) representing key points of the observed peak, i.e.\ $n=\numlist[list-final-separator = {\text{ and }}]{1;1.016;1.1}$, the last being the coordinate where diffusion recovers the value in $n=1$. These three particular configurations were further investigated separately with the aim of collecting detailed statistics at long times, with \one, \onepointosixteen and \onepointone energy packets each. The unprecedented magnitude of these simulations and the subtlety of the effects investigated required the development of dedicated software with a particular focus on numerical stability, precision and reproducibility, as described elsewhere \cite{mazzamuto2015breakdown}. In particular, dealing with more than $\approx \num{e10}$ exponentially distributed random variates generally requires the use of Pseudo Random Number Generators (PRNGs) with more than 32 random bits and sufficient floating point precision (i.e.\ long double) in order not to introduce statistically significant truncations in the drawn values, both of which are features not readily available in existing MC solutions.

\begin{figure*}
\centering
\includegraphics{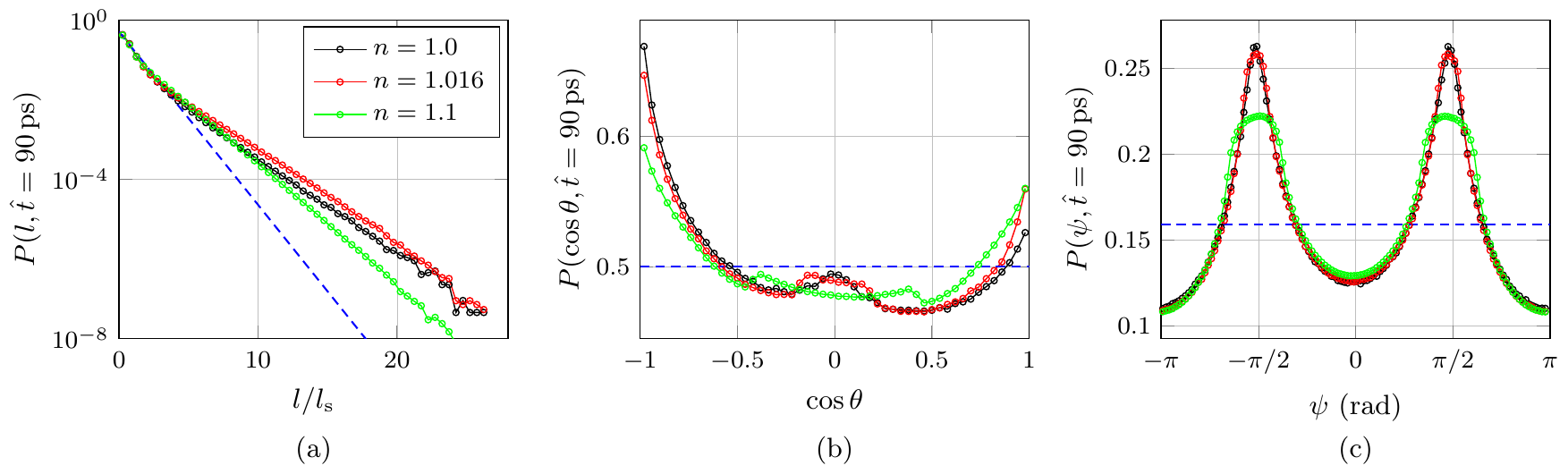}
\caption{Late-time modification of the step length and angular distributions. Three simulations with $\text{OT} = 1$, $g=0$ and $n=\numlist[list-final-separator = {\text{ and }}]{1;1.016;1.1}$ are considered. Panel (a) shows the histogram of the distribution of step lengths between consecutive scattering events performed by energy packets transmitted at $\hat{t} = \SI{90}{\pico\second}$. Also angles become unevenly sampled at late times, as shown in panels (b) and (c). Each blue dashed line represent the actual distribution fed into the simulation.}
\label{fig:actual_distributions}
\end{figure*}

As suggested by eq.~\eqref{eq:D(d)}, the most straightforward insight on the effective $D$ is obtained by directly looking at the distribution of the step lengths performed during the random walk. In principle, each energy packet is simulated according to the same step length distribution $P(l) = l_\text{s}^{-1} \exp{-l/l_\text{s}}$, which is characterized by a decaying slope and an average value of $l_\text{s}$. On the contrary, we found that there exists a clear correlation, induced by this confined geometry, between a long permanence inside the sample and an unevenly sampled step length distribution. Figure \ref{fig:actual_distributions} shows the histograms of the step lengths and scattering angles between two consecutive scattering events of energy packets transmitted at $\hat{t} = t / n_\text{in} = \SI{90}{\pico\second}$ (corresponding to a path length of $\approx 27L_0$) compared to their nominal distributions (dashed blue lines).

\looseness=-1
The step length distributions exhibit enhanced tails in this confined geometry for all the three simulated refractive index contrasts. This is consistent with a diffusion coefficient higher than the one expected by the diffusion approximation, because heavier tails will generally be associated with a greater enhancement of the second rather than the first moment (cfr.~eq.~\eqref{eq:D(d)}). The peculiar late transport regime arising in this confined geometry causes the nominal step length distribution provided by the PRNG to be sampled unevenly:
despite the fact that a long step in a very thin sample will generally cause the packet to exit the slab, those few that happen to remain inside will be able to reach long surviving times without undergoing many scattering events. Notably, in the case of refractive index contrasts very close to 1, the distribution of the step lengths features a selective enhancement of the longer values, which is slightly more marked for $n=\num{1.016}$. This might be due to the fact that, for such a small refractive index contrast, total internal reflection is already significant ($\theta_\text{c} = \ang{79.8}$). Consider as an example the case of an energy packet taking a very long step: if internal reflections are absent, extremely narrow angular conditions must hold in order for it not to exit the slab. Conversely, even a tiny contrast allows to largely relax such condition, introducing a significant increase in the survival probability of a long-stepping energy packet while only marginally affecting others. In other words there's a positive correlation between long steps and shallow incidence angles, whose effects become more apparent when such angles are the only ones undergoing total internal reflection (which also explains why the enhancement is asymmetric around $n=1$). On the other hand, with increasing contrast, more energy packets will be held inside the slab irrespective of their incidence angle (and hence of the length of their step), thus washing out the observed MSW spreading boost.

\looseness=-1
Interestingly, the sampling of the angular variables is also modified at late times, as shown in Figures \ref{fig:actual_distributions}b-c for the same set of simulations. While tracing each energy packet trajectory, the cosine of the scattering (polar) angle $\theta$ and the azimuthal angle $\psi$ are generated through the PRNG so to be uniformly distributed in $[-1,1]$ and $[0,2\pi)$, respectively. On the contrary, the $\cos \theta$ distribution at late times exhibits a back and a forward peak, with a plateau around $\cos \theta = 0$. Further insight on these features is gained by looking at the concurrent modifications in the $\psi$ statistics, which exhibit two symmetric peaks around $\pm \pi/2$. These two peaks correspond to a right/left turn in the slab plane, which helps keeping the trajectory inside the sample irrespective of the scattering angle $\theta$. The two $\cos \theta$ peaks can also be intuitively understood by considering a typical step in a very long trajectory: this will generally be a long step (i.e.~$l\gtrapprox l_\text{s}=L_0$) mostly aligned with the slab plane. As such, scattering angles close to $\theta = \ang{0}$ or \ang{180} will guarantee that the trajectory will continue within the slab irrespective of what azimuthal angle is drawn. Actually, since a typical step will not be in general perfectly parallel to the interfaces, a scattering angle of $\theta\approx\ang{180}$ should provide higher chances of staying inside the sample, hence its higher probability. Interestingly, this results in a $\cos \theta$ distribution with a slightly negative average value, which is also able to influence the effective diffusion properties exhibited by the sample.

\begin{figure}
\centering
\includegraphics{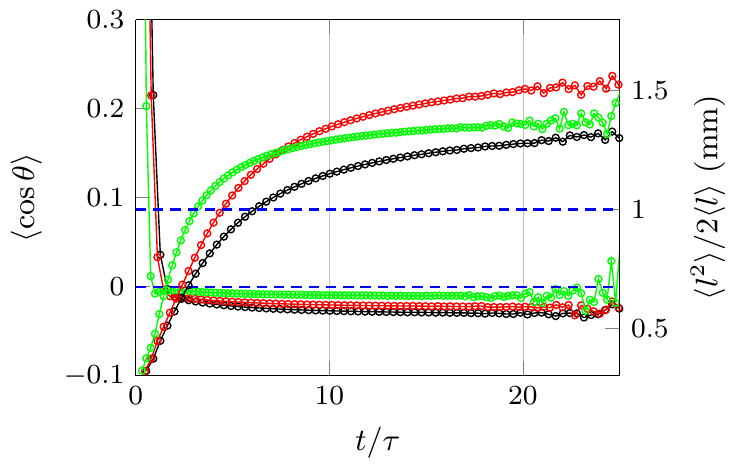}
\caption{Time evolution of the effective scattering length and effective scattering coefficient as inferred from the time-resolved effective step length and angular distributions. For each time bin, $\langle\cos\theta\rangle$ and $\langle l^2 \rangle/2\langle l \rangle$ are calculated from the retrieved time-dependent distributions, based on the trajectories of energy packets transmitted within that time bin. Blue dashed lines represent the expected values for the two distributions.}
\label{fig:effectiveD}
\end{figure}

The time dependence of both $\langle\cos\theta\rangle$ and the ratio $\langle l^2 \rangle/2\langle l \rangle$ is tracked in Figure~\ref{fig:effectiveD}, along with their nominal values (dashed blue lines). Each point collects the statistics of those energy packets that were transmitted within that time bin.
The obtained curves successfully validate our interpretation based on a random walk picture of the diffusive process as expressed by equation \eqref{eq:D(d)}, exhibiting good qualitative agreement with the mean square width expansion data. In principle, the overall observed diffusion process will be influenced by both the modified step length and angular statistics, which in the investigated configurations appear to have opposite effects. While the latter would indeed tend to slightly slow down diffusion, the predominant effect is clearly coming from the step lengths being substantially increased, leading to the observed enhanced in-plane diffusion especially for $n=\num{1.016}$.
Notably, different configurations might lead to a different overall balance between these two effects, which also appear to saturate to their respective asymptotic values with slightly different time scales, further illustrating the need for additional investigation even for the simple homogeneous and isotropic single slab model.

These results show that the currently established theoretical framework linking radiative transfer theory to diffusion needs to be further refined, especially for thin systems and more generally for confined geometries. In particular, concerning microscopic optical properties such as the scattering anisotropy or the mean free path, it seems appropriate to introduce a distinction between an \emph{intrinsic} and an \emph{effective} counterpart, where the former is the one that we are typically interested in retrieving while the latter might have a very different value and nature (e.g.~tensorial instead of scalar) depending on incidental geometric conditions. We have shown in fact that a homogeneous, isotropic slice of a certain disordered material will eventually reach an asymptotic, multiple scattering regime characterized by different (and possibly anisotropic) statistical distributions, depending on both sample size and boundary conditions.

It should be noted that such discrepancy --- which we are now able to correctly identify as the emerging of an effective transverse diffusion coefficient --- has already been reported experimentally in samples with an optical thickness as high as 8 \cite{pattelli2015direct}, suggesting that it can indeed represent an appreciable issue in a broad range of applications. It is worth stressing again the different nature of mean square width and lifetime deviations at higher optical thicknesses. Even at $\text{OT} = \num{25}$, for a slab sample with a typical glassy refractive index contrast, the diffusive prediction for the lifetime is still wrong by more than \SI{2}{\percent} while the mean square width expansion discrepancy is one order of magnitude smaller, suggesting that the lifetime deviation is due to a defective diffusion-based modeling of boundary conditions rather than to an actually different effective diffusion coefficient.

\begin{figure}
\centering
\includegraphics{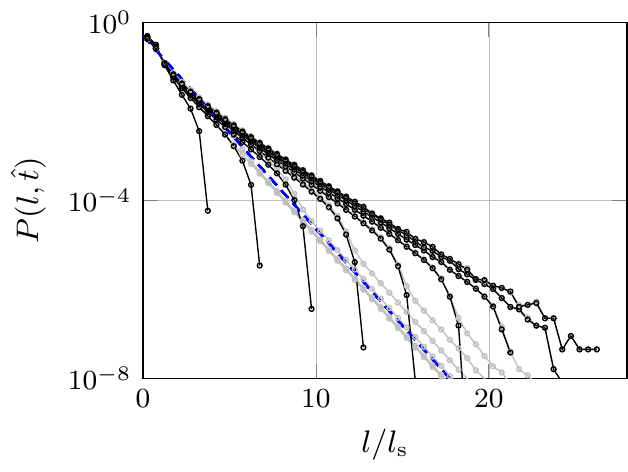}
\caption{Time evolution of the step length distribution for $n=\num{1}$ for energy packets transmitted at $\hat{t} = \SIlist[list-final-separator=\text{ and }]{10;20;30;40;50;60;70;80;90}{\pico\second}$. Black and gray curves show respectively the histograms of the lengths of the steps taken inside the sample and drawn through the PRNG. The two differ for the last step, whose length is accounted for only up to the exit surface in the first histogram. At late times the two sets of curves become indistinguishable since, as expected, the contribution of the last step to the whole trajectory becomes eventually statistically negligible.}
\label{fig:asymptotic}
\end{figure}

The asymptotic nature of the effective diffusion coefficient in a thin slab is further illustrated in Figure \ref{fig:asymptotic}, where the time evolution of the step length distribution is shown for $n=1$ (the $n=\numlist[list-pair-separator={\text{ and }}]{1.016;1.1}$ cases are analogous). The time-resolved distributions seem to converge towards a single asymptotic envelope distribution with a well defined asymptotic decay which seems to be uniquely determined by the properties of the sample. It is interesting to compare the histogram of the actual steps performed inside the sample with the histogram of the ones drawn from the PRNG. The two differ only for the last step, whose length is respectively considered either partially (up to the interface) or totally. As can be seen, only the earliest transmitted energy packets did actually sample the expected distribution evenly: in this respect, while on one hand they underwent too few scattering events and thus fall outside the diffusive approximation, on the other hand they are the only ones to migrate according to the nominal step length distribution.

\looseness=-1
Even if at an optical thickness close to \num{1} it is indeed questionable to expect diffusion predictions to hold, it is nonetheless true that we were able to follow the evolution of light deeply into the multiple scattering regime, where light is undoubtedly diffusing as the radial intensity profiles and their linear variance expansion clearly demonstrate. Following the usual approach based on the diffusion approximation, it may be argued that it could be possible to take into account all the effects discussed here by means of some refined extrapolated boundary condition. On the contrary, the optical thickness must be regarded as one of the parameters actively and independently affecting transport properties. Moreover, extrapolated boundary conditions are devised to impact and substantially correct quantities such as total transmittance which, conversely, would be hardly affected by asymptotic modifications of the effective diffusion coefficient.

The absolutely non-trivial dependence of the random walk statistics near $n=1$ should also be stressed. This is particularly relevant since index matched boundary conditions are customarily considered as a ``forgiving case'' where the diffusive approximation is expected to hold better also for thinner systems \cite{svensson2013exploiting}. While this might be true under various circumstances, from an experimental point of view it is clear that typical attempts to approximately index-match a thin sample within a reference material might more likely introduce an error rather than neutralize it, since it's the very smallest index mismatch that results in the strongest deviations from the theory.

\section{Conclusions}
We have studied the diffusive transport regime in optically thin samples, which develops at late times mainly within the slab plane and shows peculiar features especially for small refractive index contrasts. In order to explain our numerical results we performed extensive simulations of three key configurations which allowed us to inspect transport properties on exceptionally long time scales, so far unexplored. Our investigation revealed a subtle interplay occurring between the actual thickness of the slab, the refractive index contrast and the scattering anisotropy. This results in a regime that is properly diffusive but which cannot be described in terms of the standard diffusive approximation. Such interplay gives rise to a different diffusion coefficient which emerges naturally from the overall optical and geometric boundary conditions of the sample, and is univocally determined by them through yet unknown relations. In this respect, our findings recall a recently published work where it is analogously demonstrated that the link between microscopic (i.e. the scattering coefficient) and macroscopic (i.e.\ the diffusion coefficient) transport parameters remains unknown for diffusive anisotropic media \cite{alerstam2014anisotropic}, therefore reinforcing the claim for a rethink of our present picture of the diffusive modeling of a random walk-based transport problem.

As we have shown in Figures \ref{fig:effectiveD} and \ref{fig:asymptotic}, transport eventually converges to an asymptotic regime characterized by a well-defined diffusion coefficient. Interestingly, we found that because the migration of simulated energy packets is directly altered by the sample configuration, the sample itself appears generally to be less scattering than it actually is. In other words, once the diffusive regime is reached, energy packets migrate as if scatterers were further apart than they actually are, i.e.\ with an \emph{effective} transport mean free path which is greater than the one \emph{intrinsic} to the material. From this perspective, we have demonstrated that even in the presence of clear diffusion signatures (such as a steadily linear mean square width growth, see Fig.\ \ref{fig:OT1_plane}b) care has to be taken on whether it is appropriate to resort to standard diffusive approximation.

Because of the asymptotic nature of the effects here described, when studying thin samples only a small fraction of incoming light is actually subject to this effective transport mean free path. Nonetheless experimental techniques capable of detecting this discrepancy have been recently reported \cite{pattelli2015direct}. Furthermore other applications can be envisioned where multiple scattering in thin layers, even if limited to few energy packets, could play a significant role (e.g.\ random lasers).

Finally, despite our case study is in the field of light transport, random walks are extremely general models for a broad range of phenomena in complex systems, from molecular kinetics to social behaviors. Therefore our findings could have far-reaching consequences in other scenarios where tight geometric confinement holds and analogous boundary conditions can be modeled.
In this respect we demonstrated how, depending on the application, the interplay between transport properties and the environment geometry can give rise to sharp and unexpected macroscopic migration features, which is either a condition to be taken into account or one to be possibly exploited as an engineering degree of freedom to enhance transport.

\begin{acknowledgments}
This work is financially supported by the European Network of Excellence Nanophotonics for Energy Efficiency and the ERC through the Advanced Grant PhotBots (project reference 291349) funded under FP7-IDEAS-ERC. CT and GM acknowledge support from the MIUR program Atom-based Nanotechnology and from the Ente Cassa di Risparmio di Firenze with the project GRANCASSA.
\end{acknowledgments}

\bibliography{references}

\end{document}